\documentclass[12pt]{article}
\usepackage{latexsym}
\usepackage{times}
\usepackage[dvips]{graphicx}

\newcommand{\Qed}{\hspace*{1mm}\hfill$\Box$\endgraf}
\def\RR{\hbox{\sf I\kern-.14em\hbox{R}}}
\def\NN{\hbox{\sf I\kern-.13em\hbox{N}}}
\def\Min{\mathop{\rm Min}\nolimits}
\def\Max{\mathop{\rm Max}\nolimits}
\newcommand{\Units}{\mathbf{U}}
\newcommand{\Net}{\mathbf{N}}
\newcommand{\DK}[1]{\stackrel{\rightharpoonup}{\mathbf{K}}_{#1}}
\newcommand{\inv}[1]{#1^\mathrm{inv}}
\newcommand{\trecl}[1]{{#1}^\star}
\newcommand{\tracl}[1]{\overline{#1}}
\newcommand{\url}[1]{{\textbf{\texttt{\small #1}}}}

\title{Efficient Algorithms for Citation Network Analysis}
\author{Vladimir Batagelj \\
        University of Ljubljana, Department of Mathematics,\\
        Jadranska 19, 1\,111 Ljubljana, Slovenia \\
        e-mail: \texttt{vladimir.batagelj@uni-lj.si}}
\date{}

\begin{document}
\maketitle

\begin{abstract}
In the paper very efficient, linear in number of arcs, algorithms
for determining Hummon and Doreian's arc weights SPLC and SPNP in
citation network are proposed, and some theoretical properties
of these weights are presented. The nonacyclicity problem in
citation networks is discussed. An approach to identify
on the basis of arc weights an important small subnetwork is proposed
and illustrated on the citation networks of SOM (self organizing maps)
literature and US patents.
\\[4pt]
\end{abstract}

\section{Introduction}

The citation network analysis started with the
paper of Garfield et al. (1964) \cite{Gar64} in which the introduction
of the notion of citation network is attributed to Gordon Allen.
In this paper, on the example of Asimov's history of DNA \cite{Asimov},
it was shown that the analysis "\textit{demonstrated a high
degree of coincidence between an historian's account of events and the citational
relationship between these events}". An early overview of possible
applications of graph theory in citation network analysis was made
in 1965 by Garner \cite{3D}.

The next important step was made by Hummon and Doreian (1989)
\cite{HumDor89,HumDor90,HuDoFr90}. They proposed three indices
(NPPC, SPLC, SPNP) --
weights of arcs that provide us with automatic way to identify
the (most) important part of the citation network -- the main path
analysis.

In this paper we make a step further. We show how to efficiently
compute the Hummon and Doreian's weights, so that they can be used
also for analysis of very large citation networks with several
thousands of vertices. Besides this some theoretical properties
of the Hummon and Doreian's weights are presented.

The proposed methods are implemented in \texttt{\textbf{Pajek}} --
a program, for Windows (32 bit), for analysis of \emph{large networks}.
It is freely available, for noncommercial use, at its homepage
\cite{pajek}.

For basic notions of graph theory see Wilson and Watkins \cite{GT}.

\section{Citation Networks}

In a given set of units $\Units $ (articles, books, works,
\ldots) we introduce a $citing$ relation
$R \subseteq \Units  \times \Units $
 \[ u R v \equiv v \mbox{ cites } u \]
which determines a \emph{citation network} $\Net = (\Units ,R)$.

In Table~\ref{citnets} some characteristics of real life citation networks
are presented. Most of these networks were obtained from the Eugene Garfield's
collection of citation data \cite{Gar64,Gar02} produced using
\textit{\textbf{HistCite}} Software (formerly called \textit{\textbf{HistComp}}
-- \textit{comp}iled \textit{Hist}oriography program)
\cite{Gar01}. All of these networks are the result of searches in the Web of Science
and are used with the permission of ISI of Philadelphia,
\texttt{\textbf{www.isinet.com}}. These networks in \texttt{\textbf{Pajek}}'s format
are available from \texttt{\textbf{Pajek}}'s web site \cite{data}.

In Table~\ref{citnets}: $n = |\Units|$ is the number of vertices;
$m = |R|$ is the number of arcs;
$m_0$ is the number of loops;
$n_0$ is the number of isolated vertices;
$n_C$ is the size of the largest weakly connected component;
$k_C$ is the number of nontrivial weakly connected components;
$h$ is the depth of network (minimum number of levels);
$\Delta_{in}$ is the maximum input degree;
and $\Delta_{out}$ is the maximum output degree.
The last three columns contain the numbers of strongly connected components
(cyclic parts) of size 2, 3 and 4.

\begin{table}
\caption{Citation network characteristics\label{citnets}}
\begin{center}\footnotesize
\begin{tabular}{l|r|r|r|r|r|r|r|r|r|r|r|r|}
network              &  $n$ &   $m$ & $m_0$ & $n_0$ & $n_C$ & $k_C$ & $h$&$\Delta_{in}$&$\Delta_{out}$&$2$& $3$ & $4$  \\ \hline
DNA                  &   40 &    60 &     0 &     1 &    35 &     3 &  11 &          7 &          5 &   0 &   0 &  0   \\
Coupling             &  223 &   657 &     1 &     5 &   218 &     1 &  16 &         19 &        134 &   0 &   0 &  0   \\
Small world          &  396 &  1988 &     0 &   163 &   233 &     1 &  16 &         60 &        294 &   0 &   0 &  0   \\
Small \& Griffith    & 1059 &  4922 &     1 &    35 &  1024 &     1 &  28 &         89 &        232 &   2 &   0 &  0   \\
Cocitation           & 1059 &  4929 &     1 &    35 &  1024 &     1 &  28 &         90 &        232 &   2 &   0 &  0   \\
Scientometrics       & 3084 & 10416 &     1 &   355 &  2678 &    21 &  32 &        121 &        105 &   5 &   2 &  1   \\
Kroto                & 3244 & 31950 &     1 &     0 &  3244 &     1 &  32 &        166 &       3243 &   6 &   0 &  0   \\
SOM                  & 4470 & 12731 &     2 &   698 &  3704 &    27 &  24 &         51 &        735 &  11 &   0 &  0   \\
Zewail               & 6752 & 54253 &     1 &   101 &  6640 &     5 &  75 &        166 &        227 &  38 &   1 &  2   \\
Lederberg            & 8843 & 41609 &     7 &   519 &  8212 &    35 &  63 &        135 &       1098 &  54 &   4 &  0   \\
Desalination         & 8851 & 25751 &     7 &  1411 &  7143 &   115 &  27 &         73 &        137 &  12 &   0 &  1   \\
US patents     & 3774768 & 16522438 &     1 &     0 & 3764117 & 3627 & 32 &        779 &        770 &   0 &   0 &  0   \\ \hline
\end{tabular}
\end{center}
\end{table}

A citing relation is usually \emph{irreflexive},
$\forall u \in \Units  : \lnot u R u$,
and (almost) \emph{acyclic} -- no vertex is reachable from
itself by a nontrivial path, or formally
$\forall u \in \Units  \forall k \in \NN^+ : \lnot u R^k u$.
In the following we shall assume that it has this property.
We shall postpone the question how to deal with nonacyclic
citation networks till the end of the theoretical part of
the paper.

For a relation $Q \subseteq \Units  \times \Units $ we denote by
$\inv{Q}$ its \emph{inverse} relation,
$u \inv{Q} v \equiv v Q u$,
and by
 \[ Q(u) = \{ v \in \Units  : u Q v \} \]
the set of successors of unit $u \in \Units $.
If $Q$ is acyclic then also $\inv{Q}$ is acyclic.
This means that the network $\inv{\Net} = (\Units, \inv{R})$,
$u \inv{R} v \equiv  u \mbox{ cites } v$, is a network of the same
type as the original citation network $\Net = (\Units,R)$.
Therefore it is just a matter of 'taste' which relation
to select.

Let $I = \{ (u,u) : u \in \Units  \}$ be the \emph{identity} relation
on $\Units $ and $\tracl{Q} = \bigcup_{k \in \NN^+} Q^k$
the \emph{transitive closure} of relation $Q$. Then
$Q$ is acyclic iff $\tracl{Q} \cap I = \emptyset$.
The relation $\trecl{Q} = \tracl{Q} \cup I$ is the
\emph{transitive and reflexive closure} of relation $Q$.

Since the set of units $\Units $ is finite and $R$ is acyclic we
know from the theory of relations that:
\begin{itemize}
 \item The set of units $\Units $ can be \emph{topologically ordered} --
  there exists a surjective mapping (permutation) $i : \Units  \to 1 .. |\Units |$
  with the property
  \[ u R v \Rightarrow i(u) < i(v) \]
 \item Let $\Min R = \{ u \in \Units  : \inv{R}(u) = \emptyset \}$ be the set
  of \emph{minimal} elements and
   $\Max R = \{ u \in \Units  : R(u) = \emptyset \}$
  the set of \emph{maximal} elements. Then $\Min R \ne \emptyset$ and
  $\Max R \ne \emptyset$.
 \item Every unit $u \in \Units$ and every arc $(u,v) \in R$ belong
       to at least one path from  $\Min R $ to $\Max R$:\\
       $\forall u \in \Units : R^\star (u) \cap \Max R \ne \emptyset$ \\
       $\forall u \in \Units : {\inv{R}}^\star (u) \cap \Min R \ne \emptyset$
\end{itemize}

\begin{figure}
 \begin{center}
  \includegraphics[width=60mm,viewport=10 4 217 280]{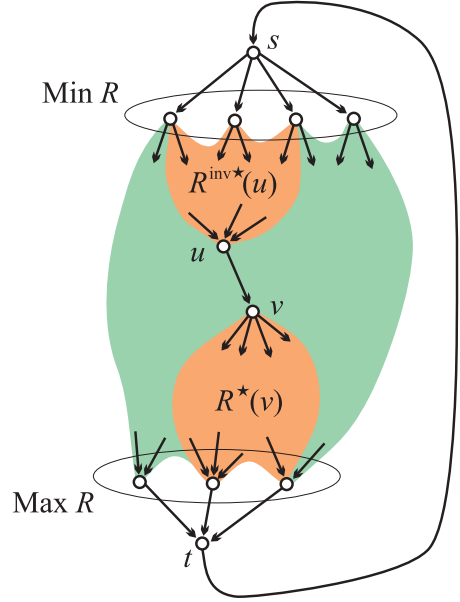}
  \caption{Citation Network in Standard Form\label{net}}
 \end{center}
\end{figure}

To simplify the presentation we transform a citation network $\Net = (\Units ,R)$ to its
\emph{standard form} $\Net' = (\Units',R')$ (see Figure~\ref{net})
by extending the set of units
$\Units ' := \Units  \cup \{ s, t \}$, $s, t \notin \Units $ with a common
\emph{source} (initial  unit) $s$ and a common \emph{sink}
(terminal unit) $t$, and by adding the corresponding arcs to relation $R$
\[ R' := R \ \cup\ \{s\} \times \Min R\ \cup\ \Max R \times \{t\}
         \ \cup\ \{ (t,s) \} \]
This eliminates problems with networks with several connected components
and/or several initial/terminal units.
In the following we shall assume that the citation network
$\Net = (\Units,R)$ is in the standard form.
Note that, to make the theory smoother, we added to $R'$ also the
'feedback' arc $(t,s)$, thus destroying its acyclicity.

\section{Analysis of Citation Networks}

An approach to the analysis of citation network is to determine
for each unit / arc its \emph{importance} or \emph{weight}. These values
are used afterward to determine the essential substructures in the
network.
In this paper we shall focus on the methods of assigning weights
$w : R \to \RR^+_0$
to arcs proposed by Hummon and Doreian \cite{HumDor89,HumDor90}:
\begin{itemize}
 \item \emph{node pair projection count}  (NPPC) method:
  $w_d(u,v) = |\trecl{\inv{R}}(u)|\cdot|\trecl{R}(v)|$
 \item \emph{search path link count}  (SPLC) method: $w_l(u,v)$ equals
  the number of "\textit{all possible search paths through the network
  emanating from an origin node}"  through the arc $(u,v) \in R$,
  \cite[p. 50]{HumDor89}.
 \item \emph{search path node pair} (SPNP) method:
  $w_p(u,v)$  "\textit{accounts for
  all connected vertex pairs along the paths through the arc $(u,v) \in R$}",
  \cite[p. 51]{HumDor89}.
\end{itemize}

\subsection{Computing NPPC weights}

To compute $w_d$ for sets of units of moderate size (up to some thousands of units)
the matrix representation of $R$ can be used and its transitive
closure computed by Roy-Warshall's algorithm \cite{algo}. The quantities
$|\trecl{R}(v)|$ and $|\trecl{\inv{R}}(u)|$ can be obtained
from closure matrix as row/column sums.
An $O(nm)$ algorithm for computing $w_d$ can be constructed using
Breath First Search from each $u \in \Units$ to determine
$|\trecl{\inv{R}}(u)|$ and $|\trecl{R}(v)|$.
Since it is of order at least $O(n^2)$ this algorithm is not suitable
for larger networks (several ten thousands of vertices).

\subsection{Search path count method}

To compute the SPLC and SPNP weights we introduce a related
\textit{search path count} (SPC) method for which the
weights $N(u,v)$, $u R v$ count the number of
different paths  from $s$ to $t$ (or from $\Min R$ to $\Max R$)
through the arc $(u,v)$.

To compute $N(u,v)$ we introduce two auxiliary quantities:
let $N^-(v)$ denotes the number of different $s$-$v$ paths,
and $N^+(v)$ denotes the number of different $v$-$t$ paths.

Every $s$-$t$ path $\pi$ containing the arc $(u,v) \in R$
can be uniquely expressed in the form
\[ \pi = \sigma \circ (u,v) \circ \tau \]
where $\sigma$ is a $s$-$u$ path and $\tau$ is a $v$-$t$ path.
Since every pair $(\sigma,\tau)$ of
$s$-$u$ / $v$-$t$ paths gives a corresponding
$s$-$t$ path  it follows:
 \[ N(u,v) = N^-(u)\cdot N^+(v), \qquad (u,v) \in R \]
where
\[
N^-(u) =
\cases{
 1  & $u = s$ \cr
 \sum_{v : v R u} N^-(v) \quad & otherwise}
\]
and
\[
N^+(u) =
\cases{
 1   & $u = t$ \cr
 \sum_{v : u R v} N^+(v) \quad & otherwise}
\]
This is the basis of an efficient algorithm for computing
the weights $N(u,v)$ --
after the topological sort of the network \cite{algo}
we can compute, using the above relations in topological order,
the weights in time of order $O(m)$.
The topological order ensures that all the quantities in
the right side expressions of the above equalities are already
computed when needed. The counters $N(u,v)$ are used as SPC
weights $w_c(u,v) = N(u,v)$.

\subsection{Computing SPLC and SPNP weights}

The description of SPLC method in \cite{HumDor89} is not very
precise. Analyzing the table of SPLC weights from
\cite[p. 50]{HumDor89} we see that we have to consider
\textbf{each} vertex as an origin of search paths.
This is equivalent to apply the SPC method on the
extended network
 $\Net_l = (\Units',R_l)$
\[ R_l := R'\ \cup\ \{ s \} \times (\Units  \setminus  \cup R(s) )  \]

It seems that there are some errors in the table of SPNP
weights in \cite[p. 51]{HumDor89}. Using the definition
of the SPNP weights we can again reduce their computation
to SPC method applied on the extended network
$\Net_p = (\Units',R_p)$
\[ R_p := R \ \cup\  \{ s \} \times \Units \  \cup \  \Units
  \times \{ t \} \ \cup\ \{ (t,s) \} \]
in which every unit $u \in U$ is additionaly linked
from the source $s$ and to the sink $t$.

\subsection{Computing the numbers of paths of length $k$}

We could use also a direct approach to determine the
weights $w_p$. Let $L^-(u)$ be the number of different
paths terminating in $u$
and $L^+(u)$ the number of different
paths originating in $u$.
Then for $uRv$ it holds $ w_p(u,v)  = L^-(u)\cdot L^+(v)$.

The procedure to determine $L^-(u)$ and $L^+(u)$ can be compactly described using two
families of polynomial generating functions\\
\[ P^-(u;x) = \sum_{k=0}^{h(u)} p^-(u,k) x^k \qquad
\mbox{and}  \qquad  P^+(u;x) = \sum_{k=0}^{h^-(u)} p^+(u,k) x^k, \quad  u \in \Units \]
where $h(u)$ is the depth of vertex $u$ in network $(\Units,R)$, and
$h^-(u)$ is the depth of vertex $u$ in network $(\Units,\inv{R})$,
The coefficient $p^-(u,k)$ counts the number of paths of length $k$ to $u$,
and $p^+(u,k)$ counts the number of paths of length $k$ from $u$.

Again, by the basic principles of combinatorics
\[
P^-(u;x) =
\cases{
 0   & $u=s$ \cr
 1 + x \cdot \sum_{v : v R u} P^-(v;x) \quad & otherwise}
\]
and
\[
P^+(u;x) =
\cases{
 0   & $u=t$ \cr
 1 + x \cdot \sum_{v : u R v} P^+(v;x) \quad & otherwise}
\]
and both families can be determined using the definitions and
computing the polynomials in the (reverse for $P^+$) topological
ordering of $\Units$. The complexity of this procedure is at most
$O(hm)$. Finally
\[  L^-(u) = P^-(u;1)     \qquad \mathrm{and} \qquad
    L^+(v) = P^+(v;1)   \]
In real life citation networks the depth $h$ is relatively small as can be seen
from the Table~\ref{citnets}.

The complexity of this approach is higher than the complexity of the
method proposed in subsection 3.3 -- but we get more detailed information about paths.
May be it would make sense to consider 'aging' of references by
$ L^-(u) = P^-(u;\alpha)$, for selected $\alpha$, $0 < \alpha \leq 1$.

\subsection{Vertex weights}

The quantities used to compute the arc weights $w$ can be used
also to define the corresponding vertex weights $t$
\begin{eqnarray*}
 t_d(u) & = & |\trecl{\inv{R}}(u)|\cdot|\trecl{R}(u)| \\
 t_c(u) & = & N^-(u)\cdot N^+(u) \\
 t_l(u) & = & N'^-(u)\cdot N'^+(u) \\
 t_p(u) & = & L^-(u)\cdot L^+(u)
\end{eqnarray*}
They are counting the number of paths of selected type through
the vertex $u$.

\subsection{Implementation details}

In our first implementation of the SPNP method the values of
$L^-(u)$ and $L^+(u)$ for some large networks (Zewail and Lederberg)
exceeded the range of Delphi's \texttt{LargeInt} (20 decimal places).
We decided to use the \texttt{Extended} real numbers
(range $= 3.6 \times 10^{-4951}\ ..\ 1.1 \times 10^{4932}$,
19-20 significant digits) for counters. This
range is safe also for very large citation networks.

To see this, let us denote $N^*(k) = \max_{u: h(u)=k} N^-(u)$.
Note that $h(s) = 0$ and $u R v \Rightarrow h(u) < h(v)$.
Let $u^* \in \Units$ be a unit on which the maximum is attained
$N^*(k) = N^-(u^*)$. Then
\begin{eqnarray*}
 N^*(k) & = & \sum_{v:v R u^*} N^-(v) \leq  \sum_{v:v R u^*} N^*(h(v)) \leq \sum_{v:v R u^*} N^*(k-1) = \\
   & = & \deg_{in}(u^*) \cdot N^*(k-1) \leq \Delta_{in}(k) \cdot N^*(k-1)
\end{eqnarray*}
where $\Delta_{in}(k)$ is the maximal input degree at depth $k$. Therefore
$N^*(h) \leq \prod_{k=1}^h \Delta_{in}(k) \leq  \Delta_{in}^h$. A similar inequality
holds also for $N^+(u)$. From both it follows
\[ N(u,v) \leq \Delta_{in}^{h(u)} \cdot \Delta_{out}^{h^-(v)} \leq \Delta^{H-1} \]
where $H = h(t)$ and $\Delta = \max(\Delta_{in}, \Delta_{out})$.
Therefore  for $H \leq 1000$ and $\Delta \leq 10000$ we get
$N(u,v)  \leq \Delta^{H-1} \leq 10^{4000}$ which is still in the range of
 \texttt{Extended} reals. Note also that in the derivation of this inequality
we were very generous -- in real-life networks $N(u,v)$ will be much smaller
than $\Delta^{H-1}$.

Very large/small numbers that result as weights in large networks are
not easy to use. One possibility to overcome this problem is to use the
logarithms of the obtained weights -- logarithmic transformation
is monotone and therefore preserve the ordering of weights (importance
of vertices and arcs). The transformed values are also more convenient
for visualization with line thickness of arcs.

\section{Properties of weights}

\subsection{General properties of weights}

Directly from the definitions of weights we get
\[  w_k(u,v;R) = w_k(v,u;\inv{R}), \qquad k=d,c,p \]
and
\[ w_c(u,v) \leq w_l(u,v) \leq  w_p(u,v) \]

Let $\Net_A = (\Units_A, R_A)$ and $\Net_B = (\Units_B, R_B)$,
$\Units_A \cap \Units_B = \emptyset$ be two citation networks,
and $\Net_1 = (\Units'_A, R'_A)$
and $\Net_2 = ((\Units_A \cup \Units_B)', (R_A \cup R_B)')$
the corresponding standardized networks of the first network
and of the union of both networks. Then it holds for all
$u,v \in \Units_A$ and for all $p,q \in R_A$
\[  \frac{t_k^{(1)}(u)}{t_k^{(1)}(v)} = \frac{t_k^{(2)}(u)}{t_k^{(2)}(v)}, \qquad
\mbox{and} \qquad
    \frac{w_k^{(1)}(p)}{w_k^{(1)}(q)} = \frac{w_k^{(2)}(p)}{w_k^{(2)}(q)}, \qquad k=d,c,l,p \]
where $t^{(1)}$ and $w^{(1)}$ is a weight on network $\Net_1$, and
$t^{(2)}$ and $w^{(2)}$ is a weight on network $\Net_2$.
This means that adding or removing components in a network
do not change the ratios (ordering) of the weights inside components.

Let $\Net_1 = (\Units,R_1)$ and $\Net_2 = (\Units,R_2)$ be two citation networks over the same
set of units $\Units$ and $R_1 \subseteq R_2$ then
\[ w_k(u,v;R_1)  \leq w_k(u,v;R_2), \qquad k=d,c,p  \]

\subsection{NPPC weights}

In an acyclic network for every arc $(u,v) \in R$ hold
\[ \trecl{\inv{R}}(u) \cap \trecl{R}(v) = \emptyset \quad \mathrm{and} \quad
   \trecl{\inv{R}}(u) \cup \trecl{R}(v) \subseteq \Units \]
therefore $|\trecl{\inv{R}}(u)| + |\trecl{R}(v)| \leq n$ and,
using the inequality $\sqrt{ab} \leq \frac{1}{2} (a+b)$, also
\[ w_d(u,v) = |\trecl{\inv{R}}(u)| \cdot |\trecl{R}(v)| \leq \frac{1}{4} n^2 \]

Close to the source or sink the weights $w_d$ are small,
since the sets $\trecl{R}(u)$ (and $\trecl{\inv{R}}(u)$) are monotonic
along the paths in a sense
\[ u \tracl{R} v \Rightarrow \trecl{R}(u) \subset \trecl{R}(v) \]
The weights $w_d$ are larger in the 'middle' of the network.

A more uniform (but less sensitive)
weight would be $w_s(u,v) = |\trecl{\inv{R}}(u)| + |\trecl{R}(v)|$
or in the normalized form $w'_s(u,v) = \frac{1}{n} w_s(u,v)$.

\subsection{SPC weights}

For the flow $N(u,v)$ the \emph{Kirchoff's node law} holds:

For every node $v$ in a citation network in standard
form it holds
\[ \mbox{incoming flow} = \mbox{outgoing flow} = t_c(v)\]

\noindent\textbf{Proof:}
\[ \sum_{x:xRv} N(x,v) = \sum_{x:xRv} N^-(x)\cdot N^+(v) =
   (\sum_{x:xRv} N^-(x))\cdot N^+(v) = N^-(v)\cdot N^+(v) \]
\[ \sum_{y:vRy} N(v,y) = \sum_{y:vRy} N^-(v)\cdot N^+(y) =
   N^-(v)\cdot\sum_{y:vRy} N^+(y) = N^-(v)\cdot N^+(v) \]
\Qed

From the Kirchoff's node law it follows that
the \emph{total flow} through the citation network equals
$N(t,s)$. This gives us a natural way to normalize the weights
 \[ w(u,v) = \frac{N(u,v)}{N(t,s)} \quad \Rightarrow \quad
    0 \leq w(u,v) \leq 1 \]
If $C$ is a minimal arc-cut-set
 \[ \sum_{(u,v) \in C} w(u,v) = 1 \]

Let $\DK{n} = \{ (u,v): u,v \in 1..n \land u < v \}$ be the
complete acyclic directed graph on $n$ vertices then the value
of $N(u,v;\DK{n})$ is maximum over all citation networks on $n$
units. It is easy to verify that
 \[ N(1,n;\DK{n}) = 2^{n-2} \]
and in general
 \[ N(i,j;\DK{n}) = 2^{j-i-1}, i < j \]
From this result we see that the exhaustive search algorithm proposed
in Hummon and Doreian \cite{HumDor89,HumDor90} can require
exponential time to compute the arc weights $w$.


\section{Nonacyclic citation networks}

The problem with cycles is that if there is a cycle in a network
then there is also an infinite number of trails between some
units. There are some standard approaches to overcome the problem:
\begin{itemize}
 \item to introduce some 'aging' factor which makes the total weight of all trails
   converge to some finite value;
 \item to restrict the definition of a weight to some finite subset of
   trails -- for example paths or geodesics.
\end{itemize}
But, new problems arise: What is the right value of the 'aging' factor?
Is there an efficient algorithm to count the restricted trails?

\begin{figure}
 \begin{center}
  \includegraphics[width=140mm,viewport=0 0 376 186,clip]{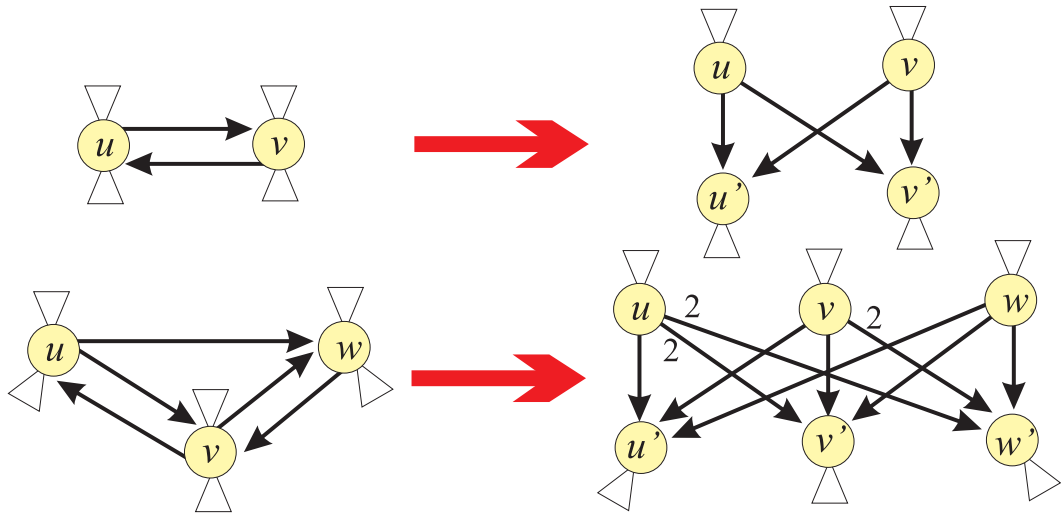}
  \caption{Preprint transformation\label{preprint}}
 \end{center}
\end{figure}

The other possibility, since a citation network is usually almost acyclic,
is to transform it into an acyclic network
\begin{itemize}
 \item by identification (shrinking) of cyclic groups (nontrivial strong
       components), or
 \item by deleting some arcs, or
 \item by transformations such as the 'preprint' transformation
       (see Figure~\ref{preprint}) which is based on the following idea:
       Each paper from a strong component is duplicated with its 'preprint'
       version. The papers inside strong component cite preprints.
\end{itemize}

Large strong components in citation network are unlikely --
their presence usually indicates an error in the data.
An exception from this rule is the
citation network of High Energy Particle Physics literature \cite{HEP}
from \textbf{arXiv}. In it different versions of the same paper
are treated as a unit. This leads to large strongly connected
components. The idea of preprint transformation can be used also in
this case to eliminate cycles.

\section{First Example: SOM citation network}

The purpose of this example is not the analysis of the selected
citation network on SOM (self-organizing maps) literature \cite{Gar02,SOM,SOMLVQ},
but to present typical steps and results in citation network analysis.
We made our analysis using program  \texttt{\textbf{Pajek}}.

First we test the network for acyclicity.
Since in the SOM network there are 11 nontrivial strong components
of size 2, see Table~\ref{citnets},
we have to transform the network into acyclic one. We decided to do
this by shrinking each component into a single vertex. This operation
produces some loops that should be removed.

Now, we can compute the citation weights. We selected the
SPC (search path count) method. It returns the following results:
the network with citation weights on arcs,  the main path
network and the vector with vertex weights.

\begin{figure}[!]
 \begin{center}
  \includegraphics[height=140mm,viewport=150 20 710 775,clip=]{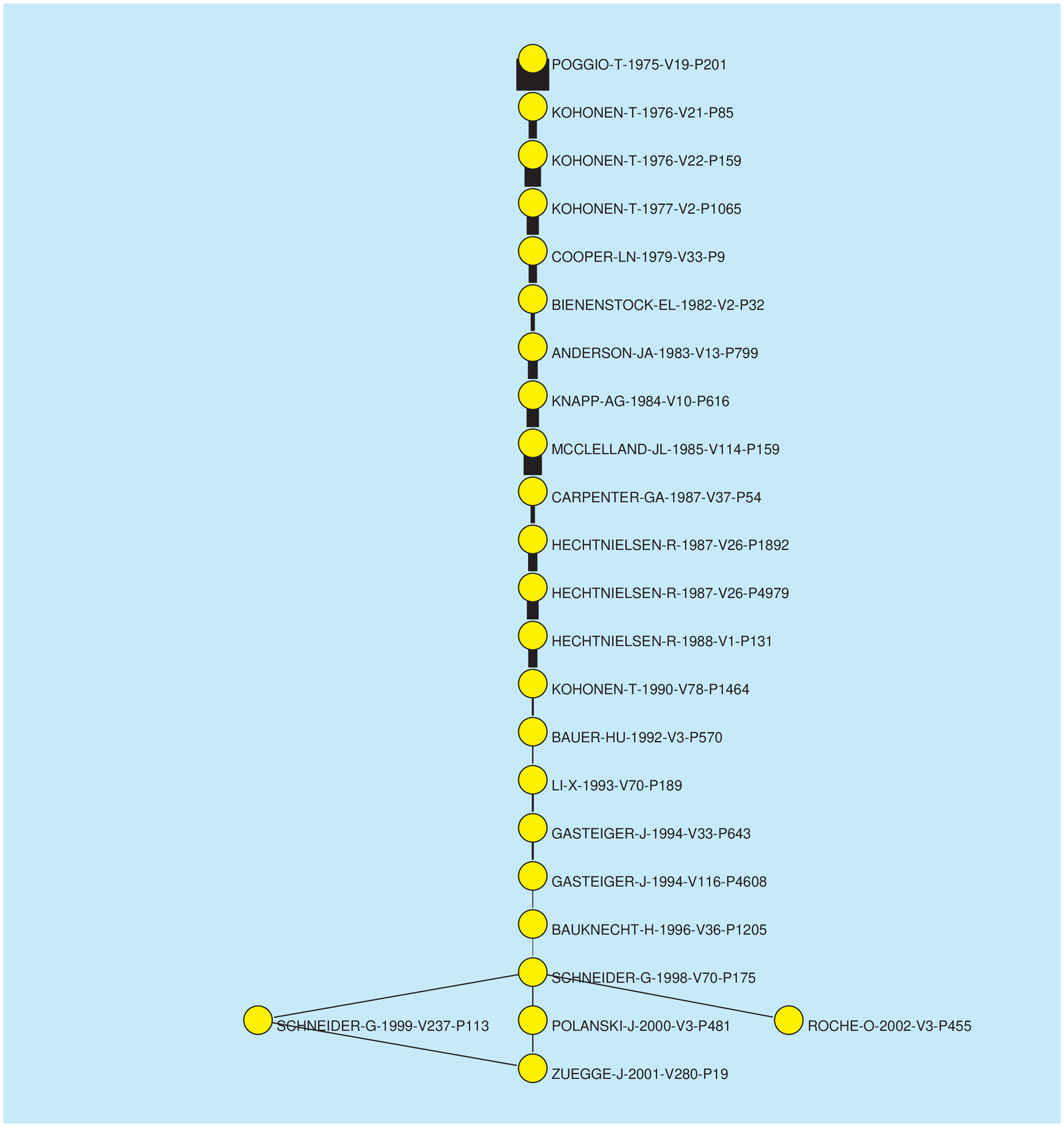}\quad
  \includegraphics[height=140mm,viewport=320 20 580 775,clip=]{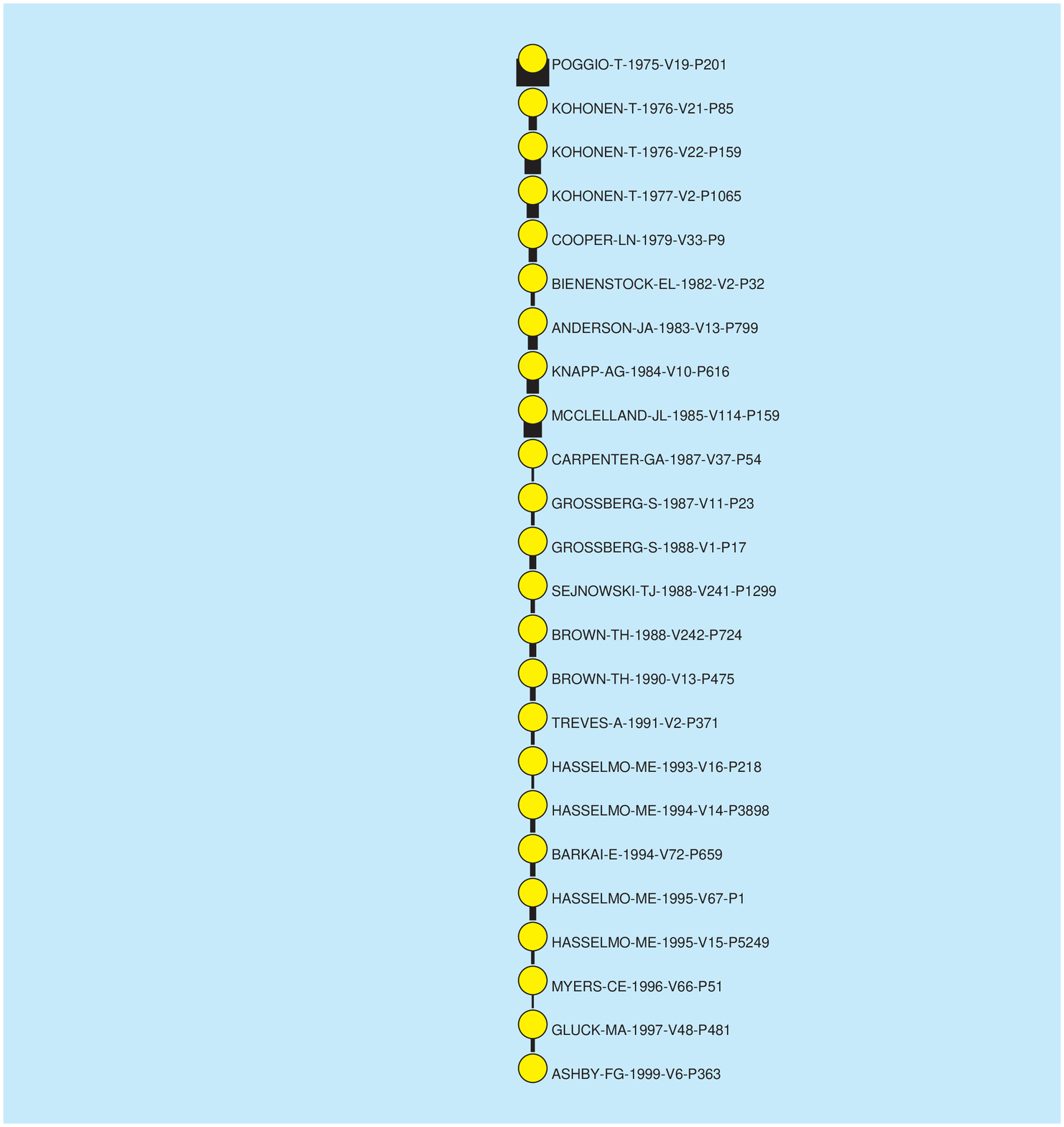}
  \caption{Main path and CPM path in SOM network with SPC weights\label{main}}
 \end{center}
\end{figure}

In a citation network, a \emph{main path}  (sub)network is
constructed starting from the source vertex
and selecting at each step in the end vertex/vertices the arc(s)
with the highest weight, until a sink vertex is reached.

Another possibility is to apply  on the network $\Net = (\Units,R,w)$
the critical path method (CPM) from operations research.

First we draw the main path network. The arc weights are represented
by the thickness of arcs. To produce a nice picture
of it we apply the Pajek's macro \texttt{Layers} which contains a
sequence of operations for determining a layered layout of an
acyclic network (used also in analysis of genealogies represented
by p-graphs). Some experiments with settings of
different options are needed to obtain a right picture,
see left part of Figure~\ref{main}. In its right part the
CPM path is presented.

We see that the upper parts of both paths are identical, but
they differ in the continuation. The arcs in the CPM path are
thicker.

We could display also the complete SOM network using
essentially the same procedure as for the displaying of
main path. But the obtained picture would be too complicated
(too many vertices and arcs). We have to identify some
simpler and important subnetworks inside it.

Inspecting the distribution of values of weights on arcs (lines)
we select a threshold 0.007 and determine the corresponding
\emph{arc-cut} -- delete all arcs with weights
lower than selected threshold and afterwards  delete also all
isolated vertices (degree $= 0$).

Now, we are ready to draw the reduced network. We first produce
an automatic layout.
We notice some small unimportant components. We preserve only
the large main component, draw it and improve the obtained layout
manually. To preserve the level structure we use the option
that allows only the horizontal movement of vertices.

\begin{figure}[!]
 \begin{center}
  \includegraphics[width=160mm,viewport=70 15 795 765,clip=]{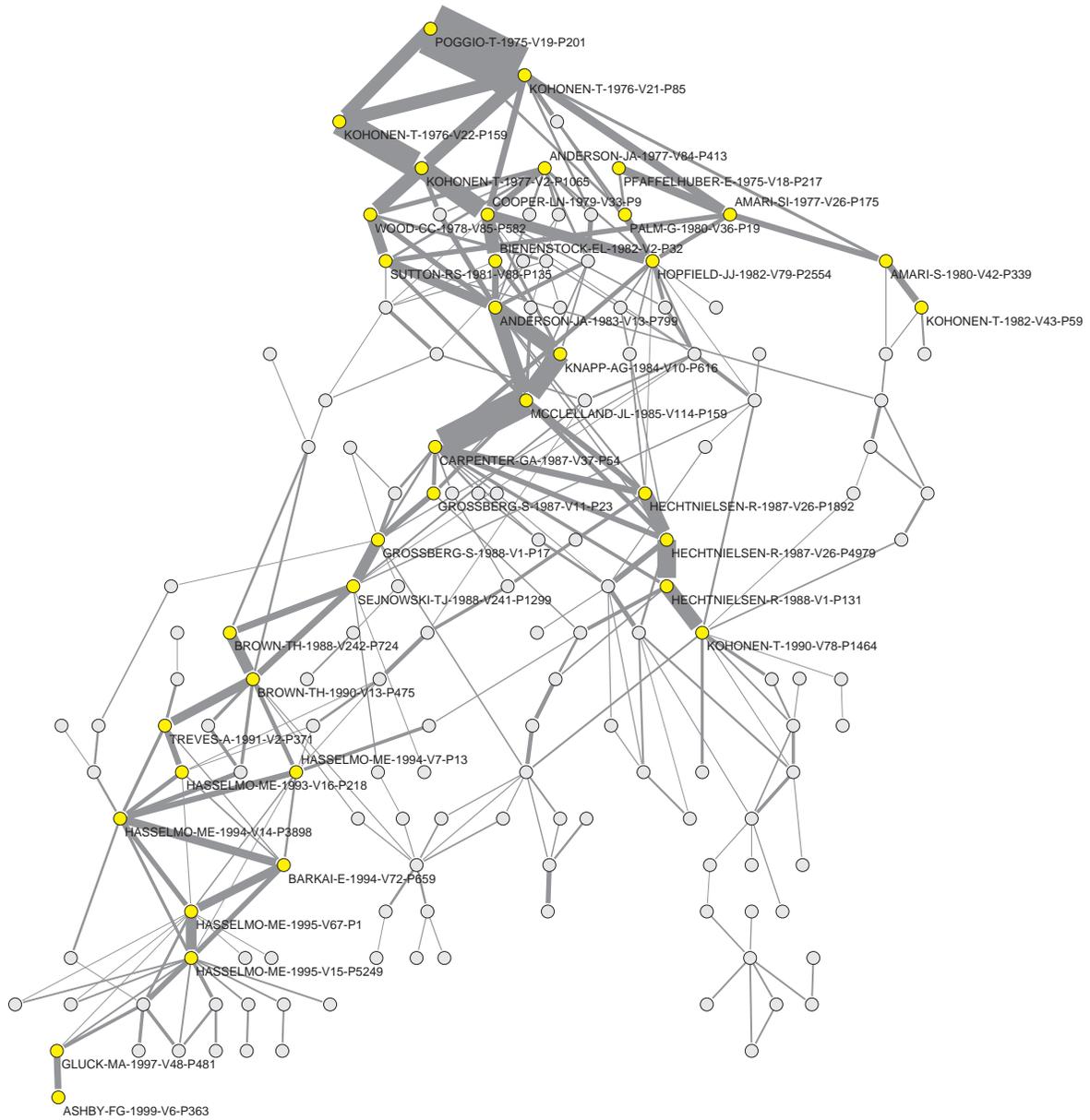}
  \caption{Main subnetwork at level 0.007\label{maina}}
 \end{center}
\end{figure}

Finally we label the 'most important vertices'
with their labels. A vertex is considered important if it is an
endpoint of an arc with the weight above the selected
threshold (in our case 0.05).

The obtained picture of SOM 'main subnetwork'
is presented in Figure~\ref{maina}.
We see that the SOM field evolved in
two main branches. From CARPENTER-1987 the strongest (main path)
arc is leading to the right branch that after some steps disappears.
The left, more vital branch is detected by the CPM path.
Further investigation of this is left
to the readers with additional knowledge about the SOM field.

\begin{table}[!]
\caption{15 Hubs and Authorities \label{huau}}
\begin{center}\small
\begin{tabular}{r|l|l|l|l|}
    Rank  & $h$       &   Hub Id                        &       $a$       &   Authority Id                          \\ \hline
       1  & 0.06442   &   CLARK-JW-1991-V36-P1259       &       0.85214   &   HOPFIELD-JJ-1982-V79-P2554  \\
       2  & 0.06366   &   \#GARDNER-E-1988-V21-P257     &       0.33427   &   KOHONEN-T-1982-V43-P59        \\
       3  & 0.05794   &   HUANG-SH-1994-V17-P212        &       0.14531   &   KOHONEN-T-1990-V78-P1464    \\
       4  & 0.05721   &   GULATI-S-1991-V33-P173        &       0.12398   &   CARPENTER-GA-1987-V37-P54   \\
       5  & 0.05513   &   SHUBNIKOV-EI-1997-V64-P989    &       0.10376   &   \#GARDNER-E-1988-V21-P257    \\
       6  & 0.05496   &   MARSHALL-JA-1995-V8-P335      &       0.09353   &   HOPFIELD-JJ-1986-V233-P625  \\
       7  & 0.05488   &   VEMURI-V-1993-V36-P203        &       0.07882   &   MCELIECE-RJ-1987-V33-P461   \\
       8  & 0.05409   &   CHENG-B-1994-V9-P2            &       0.07656   &   KOHONEN-T-1988-V1-P3          \\
       9  & 0.05360   &   BUSCEMA-M-1998-V33-P17        &       0.07372   &   RUMELHART-DE-1985-V9-P75    \\
      10  & 0.05258   &   XU-L-1993-V6-P627             &       0.07271   &   KOSKO-B-1988-V18-P49          \\
      11  & 0.05249   &   WELLS-DM-1998-V41-P173        &       0.07246   &   ANDERSON-JA-1977-V84-P413   \\
      12  & 0.05233   &   SCHYNS-PG-1991-V15-P461       &       0.07033   &   AMARI-SI-1977-V26-P175        \\
      13  & 0.05173   &   SMITH-KA-1999-V11-P15         &       0.06709   &   KOSKO-B-1987-V26-P4947        \\
      14  & 0.05149   &   BONABEAU-E-1998-V9-P1107      &       0.05802   &   PERSONNAZ-L-1985-V46-PL359  \\
      15  & 0.05126   &   KOHONEN-T-1990-V78-P1464      &       0.05702   &   GROSSBERG-S-1987-V11-P23    \\ \hline
\end{tabular}
\end{center}
\end{table}

As a complementary information we can determine
Kleinberg's hubs and authorities vertex weights \cite{ha}.
Papers that are cited by many other papers are called authorities;
papers that cite many other documents are
called hubs.
Good authorities are those that are cited by good hubs
and good hubs cite good authorities. The 15 highest
ranked hubs and authorities are presented in Table~\ref{huau}.
We see that the main authorities are located in eighties
and the main hubs in nineties.
Note that, since we are using the relation
$u R v \equiv u \mbox{ is cited by } v$, we have to
interchange the roles of hubs and authorities produced by
\texttt{\textbf{Pajek}}.

An elaboration of the hubs and authorities approach to the analysis
of citation networks complemented with visualization can be found in
Brandes and Willhalm (2002) \cite{BW}.

\section{Second Example: US patents}

The network of US patents from 1963 to 1999 \cite{patents} is an
example of very large network (3774768 vertices and 16522438 arcs)
that, using some special options in \texttt{\textbf{Pajek}},
can still be analyzed on PC with at least 1 G memory.
The SPC weights are determined in a range of 1 minute.
This shows that the proposed approach can be used also for
very large networks.

\begin{figure}[!]
 \begin{center}
  \includegraphics[height=175mm,viewport=65 20 375 635,clip=]{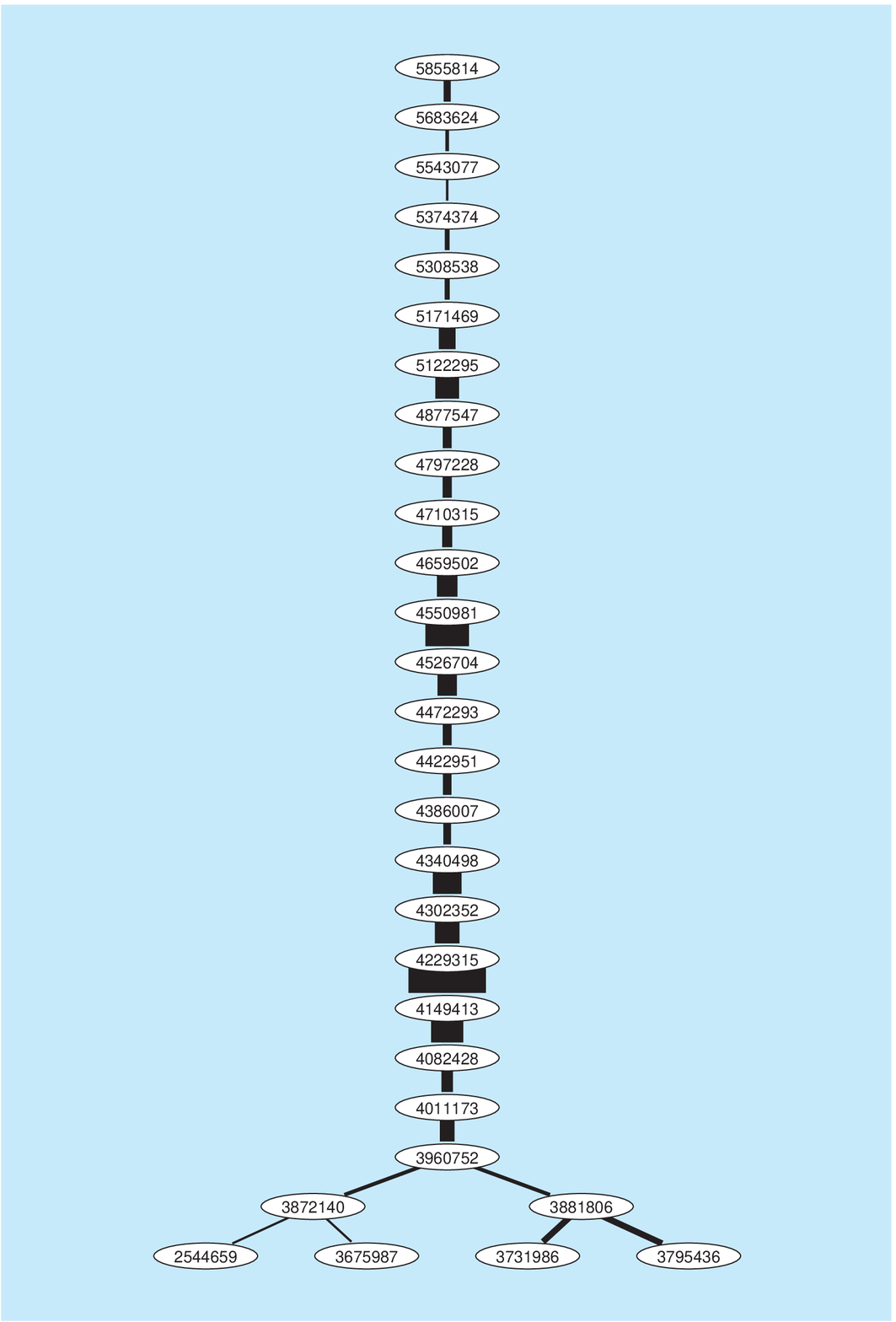}\quad
  \includegraphics[height=175mm,viewport=0 20 280 785,clip=]{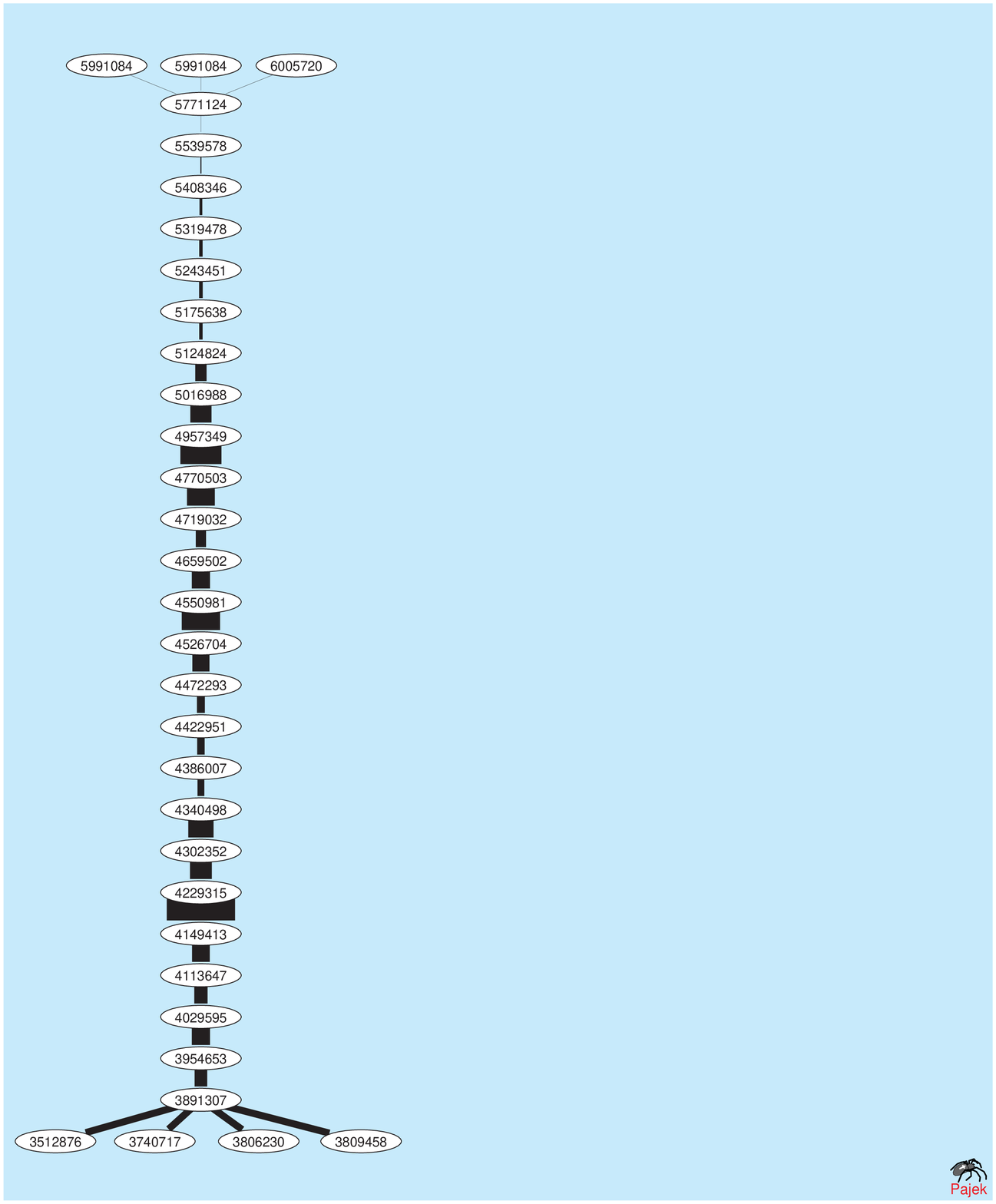}
  \caption{Main path and CPM path subnetwork of Patents\label{mainpat}}
 \end{center}
\end{figure}

The obtained main path and CPM path are presented in Figure~\ref{mainpat}.
Collecting from the
\textbf{\textit{United States Patent and Trademark Office}} \cite{uspto}
the basic data about the patents from both paths, see
Table~\ref{patinfo}-\ref{patinfoD}, we see that they deal with
'liquid crystal displays'.

\begin{table}
\caption{Patents on the liquid-crystal display\label{patinfo}}
\begin{center}
\renewcommand{\arraystretch}{0.83}
\begin{tabular}{|r|r|l|}
\hline
patent  & date         & author(s) and title \\
\hline
2544659 & Mar 13, 1951 & Dreyer.
        Dichroic light-polarizing sheet and the like and the\\
    & & formation and use thereof\\

2682562 &  Jun 29, 1954  &   Wender, et al.
        Reduction of aromatic carbinols\\

3322485 &  May 30, 1967  &   Williams.
        Electro-optical elements utilazing an organic\\
    & & nematic compound\\

3512876 &  May 19, 1970  & Marks.
        Dipolar electro-optic structures\\

3636168 &  Jan 18, 1972  &   Josephson.
        Preparation of polynuclear aromatic compounds\\

3666948 &  May 30, 1972  &   Mechlowitz, et al.
        Liquid crystal termal imaging system\\
    & & having an undisturbed image on a disturbed background\\

3675987 & Jul 11, 1972 &  Rafuse.
        Liquid crystal compositions and devices \\

3691755 &  Sep 19, 1972  &   Girard.
        Clock with digital display\\

3697150 &  Oct 10, 1972  &   Wysochi.
        Electro-optic systems in which an electrophoretic-\\
    & & like or dipolar material is dispersed throughout a liquid\\
    & & crystal to reduce the turn-off time\\

3731986 & May  8, 1973 &  Fergason.
        Display devices utilizing liquid crystal light\\
    & & modulation \\

3740717 &  Jun 19, 1973 &  Huener, et al.
           Liquid crystal display \\

3767289 &  Oct 23, 1973  &   Aviram, et al.
        Class of stable trans-stilbene compounds,\\
    & & some displaying nematic mesophases at or near room\\
    & & temperature and others in a range up to 100$^\circ$C\\

3773747 &  Nov 20, 1973  &   Steinstrasser.
        Substituted azoxy benzene compounds\\

3795436 & Mar  5, 1974 &  Boller, et al.
        Nematogenic material which exhibit the Kerr\\
    & & effect at isotropic temperatures \\

3796479 &  Mar 12, 1974  &   Helfrich, et al.
        Electro-optical light-modulation cell\\
    & & utilizing a nematogenic material which exhibits the Kerr\\
    & & effect at isotropic temperatures\\

3806230 &  Apr 23, 1974  & Haas.
        Liquid crystal imaging system having optical storage\\
    & & capabilities\\

3809458 &  May  7, 1974 &  Huener, et al.
        Liquid crystal display\\

3872140 & Mar 18, 1975 &  Klanderman, et al.
        Liquid crystalline compositions and\\
    & & method \\

3876286 & Apr  8, 1975 & Deutscher, et al.
        Use of nematic liquid crystalline substances\\

3881806 & May  6, 1975 &  Suzuki.
        Electro-optical display device \\

3891307 & Jun 24, 1975 & Tsukamoto, et al.
        Phase control of the voltages applied to\\
    & & opposite electrodes for a cholesteric to nematic phase\\
    & & transition display \\

3947375 &  Mar 30, 1976 &  Gray, et al.
        Liquid crystal materials and devices \\

3954653 &  May  4, 1976 &  Yamazaki.
        Liquid crystal composition having high dielectric\\
    & & anisotropy and display device incorporating same \\

3960752 & Jun  1, 1976 &  Klanderman, et al.
        Liquid crystal compositions \\

3975286 &  Aug 17, 1976 &  Oh.
        Low voltage actuated field effect liquid crystals\\
    & & compositions and method of synthesis \\

4000084 &  Dec 28, 1976 &  Hsieh, et al.
        Liquid crystal mixtures for electro-optical\\
    & & display devices \\

4011173 & Mar  8, 1977 &  Steinstrasser.
        Modified nematic mixtures with\\
    & & positive dielectric anisotropy \\

4013582 &  Mar 22, 1977 &  Gavrilovic.
        Liquid crystal compounds and electro-optic\\
    & & devices incorporating them \\

4017416 &  Apr 12, 1977 &  Inukai, et al.
        P-cyanophenyl 4-alkyl-4'-biphenylcarboxylate,\\
    & & method for preparing same and liquid crystal compositions\\
    & & using same \\

\hline
\end{tabular}
\end{center}
\end{table}

\begin{table}
\caption{Patents on the  liquid-crystal display\label{patinfoB}}
\begin{center}
\renewcommand{\arraystretch}{0.83}
\begin{tabular}{|r|r|l|}
\hline
patent  & date         & author(s) and title \\
\hline

4029595 &  Jun 14, 1977 &  Ross, et al.
        Novel liquid crystal compounds and electro-optic\\
    & & devices incorporating them \\

4032470 &  Jun 28, 1977 &  Bloom, et al.
        Electro-optic device \\

4077260 &  Mar  7, 1978 &  Gray, et al.
        Optically active cyano-biphenyl compounds and\\
    & & liquid crystal materials containing them \\

4082428 & Apr  4, 1978 &  Hsu.
        Liquid crystal composition and method \\

4083797 &  Apr 11, 1978 &  Oh.
        Nematic liquid crystal compositions \\

4113647 &  Sep 12, 1978 &  Coates, et al.
        Liquid crystalline materials \\

4118335 &  Oct  3, 1978 &  Krause, et al.
        Liquid crystalline materials of reduced viscosity \\

4130502 &  Dec 19, 1978 &  Eidenschink, et al.
        Liquid crystalline cyclohexane derivatives \\

4149413 & Apr 17, 1979 &  Gray, et al.
        Optically active liquid crystal mixtures and\\
    & & liquid crystal devices containing them \\

4154697 &   May 15, 1979 &   Eidenschink, et al.
        Liquid crystalline hexahydroterphenyl\\
    & & derivatives \\

4195916 &  Apr  1, 1980 &  Coates, et al.
        Liquid crystal compounds \\

4198130 &  Apr 15, 1980 &  Boller, et al.
        Liquid crystal mixtures \\

4202791 &  May 13, 1980 &  Sato, et al.
        Nematic liquid crystalline materials \\

4229315 & Oct 21, 1980 &  Krause, et al.
        Liquid crystalline cyclohexane derivatives \\

4261652 &  Apr 14, 1981 &  Gray, et al.
        Liquid crystal compounds and materials and \\
    & & devices containing them \\

4290905 &  Sep 22, 1981 &  Kanbe.
        Ester compound \\

4293434 &  Oct  6, 1981 &  Deutscher, et al.
        Liquid crystal compounds \\

4302352 & Nov 24, 1981 &  Eidenschink, et al.
        Fluorophenylcyclohexanes, the preparation\\
    & & thereof and their use as components of liquid crystal dielectrics \\

4330426 &  May 18, 1982 &  Eidenschink, et al.
        Cyclohexylbiphenyls, their preparation and\\
    & & use in dielectrics and electrooptical display elements \\

4340498 & Jul 20, 1982 &  Sugimori.
        Halogenated ester derivatives \\

4349452 &  Sep 14, 1982 &  Osman, et al.
        Cyclohexylcyclohexanoates\\

4357078 &  Nov  2, 1982 &  Carr, et al.
        Liquid crystal compounds containing an alicyclic \\
    & & ring and exhibiting a low dielectric anisotropy and liquid\\
    & & crystal materials and devices incorporating such compounds \\

4361494 &  Nov 30, 1982 &  Osman, et al.
        Anisotropic cyclohexyl cyclohexylmethyl ethers \\

4368135 &  Jan 11, 1983 &  Osman.
        Anisotropic compounds with negative or positive\\
    & & DC-anisotropy and low optical anisotropy \\

4386007 & May 31, 1983 &  Krause, et al.
        Liquid crystalline naphthalene derivatives \\

4387038 &  Jun  7, 1983 &  Fukui, et al.
        4-(Trans-4'-alkylcyclohexyl) benzoic acid \\
    & & 4'"-cyano-4"-biphenylyl esters \\

4387039 &  Jun  7, 1983 &  Sugimori, et al.
        Trans-4-(trans-4'-alkylcyclohexyl)-cyclohexane\\
    & & carboxylic acid 4'"-cyanobiphenyl ester \\

4400293 &  Aug 23, 1983 &  Romer, et al.
        Liquid crystalline cyclohexylphenyl derivatives \\

4415470 &  Nov 15, 1983 &  Eidenschink, et al.
        Liquid crystalline fluorine-containing \\
    & & cyclohexylbiphenyls and dielectrics and electro-optical display\\
    & & elements based thereon \\

4419263 &  Dec  6, 1983 &  Praefcke, et al.
        Liquid crystalline cyclohexylcarbonitrile\\
    & & derivatives \\

4422951 & Dec 27, 1983 &  Sugimori, et al.
        Liquid crystal benzene derivatives \\

4455443 &  Jun 19, 1984 &  Takatsu, et al.
        Nematic halogen Compound \\

4456712 &  Jun 26, 1984 &  Christie, et al.
        Bismaleimide triazine composition \\

4460770 &  Jul 17, 1984 &  Petrzilka, et al.
        Liquid crystal mixture \\

4472293 & Sep 18, 1984 &  Sugimori, et al.
        High temperature liquid crystal substances of\\
    & & four rings and liquid crystal compositions containing the same \\

\hline
\end{tabular}
\end{center}
\end{table}

\begin{table}
\caption{Patents on the liquid-crystal display\label{patinfoC}}
\begin{center}
\renewcommand{\arraystretch}{0.83}
\begin{tabular}{|r|r|l|}
\hline
patent  & date         & author(s) and title \\
\hline

4472592 &  Sep 18, 1984 &  Takatsu, et al.
        Nematic liquid crystalline compounds \\

4480117 &  Oct 30, 1984 &  Takatsu, et al.
        Nematic liquid crystalline compounds \\

4502974 &  Mar  5, 1985 &  Sugimori, et al.
        High temperature liquid-crystalline ester\\
    & & compounds \\

4510069 &  Apr  9, 1985 &  Eidenschink, et al.
        Cyclohexane derivatives \\

4514044 &  Apr 30, 1985 &  Gunjima, et al.
        1-(Trans-4-alkylcyclohexyl)-2-(trans-4'-(p-sub\-\\
    & & stituted phenyl) cyclohexyl)ethane and liquid crystal mixture \\

4526704 & Jul  2, 1985 &  Petrzilka, et al.
        Multiring liquid crystal esters \\

4550981 & Nov  5, 1985 &  Petrzilka, et al.
        Liquid crystalline esters and mixtures \\

4558151 &  Dec 10, 1985 &  Takatsu, et al.
        Nematic liquid crystalline compounds \\

4583826 &  Apr 22, 1986 &  Petrzilka, et al.
        Phenylethanes \\

4621901 &  Nov 11, 1986 &  Petrzilka, et al.
        Novel liquid crystal mixtures \\

4630896 &  Dec 23, 1986 &  Petrzilka, et al.
        Benzonitriles \\

4657695 &  Apr 14, 1987 &  Saito, et al.
        Substituted pyridazines \\

4659502 & Apr 21, 1987 &  Fearon, et al.
        Ethane derivatives \\

4695131 &  Sep 22, 1987 &  Balkwill, et al.
        Disubstituted ethanes and their use in liquid\\
    & & crystal materials and devices \\

4704227 &  Nov  3, 1987 &  Krause, et al.
        Liquid crystal compounds \\

4709030 &  Nov 24, 1987 &  Petrzilka, et al.
        Novel liquid crystal mixtures \\

4710315 & Dec  1, 1987 &  Schad, et al.
        Anisotropic compounds and liquid crystal\\
    & & mixtures therewith \\

4713197 &  Dec 15, 1987 &  Eidenschink, et al.
        Nitrogen-containing heterocyclic compounds \\

4719032 &  Jan 12, 1988 &  Wachtler, et al.
        Cyclohexane derivatives \\

4721367 &  Jan 26, 1988 &  Yoshinaga, et al.
        Liquid crystal device \\

4752414 &  Jun 21, 1988 &  Eidenschink, et al.
        Nitrogen-containing heterocyclic compounds \\

4770503 &  Sep 13, 1988 &  Buchecker, et al.
        Liquid crystalline compounds \\

4795579 &  Jan  3, 1989 &  Vauchier, et al.
        2,2'-difluoro-4-alkoxy-4'-hydroxydiphenyls and\\
    & & their derivatives, their production process and\\
    & & their use in liquid crystal display devices \\

4797228 & Jan 10, 1989 &  Goto, et al.
        Cyclohexane derivative and liquid crystal\\
    & & composition containing same \\

4820839 &  Apr 11, 1989 &  Krause, et al.
        Nitrogen-containing heterocyclic esters \\

4832462 &  May 23, 1989 &  Clark, et al.
        Liquid crystal devices \\

4877547 & Oct 31, 1989 &  Weber, et al.
        Liquid crystal display element \\

4957349 &  Sep 18, 1990 &  Clerc, et al.
        Active matrix screen for the color display of\\
    & & television pictures, control system and process for producing\\
    & & said screen \\

5016988 &  May 21, 1991 &  Iimura.
        Liquid crystal display device with a birefringent\\
    & & compensator \\

5016989 &  May 21, 1991 &  Okada.
        Liquid crystal element with improved contrast and\\
    & & brightness \\

5122295 & Jun 16, 1992 &  Weber, et al.
        Matrix liquid crystal display \\

5124824 &  Jun 23, 1992 &  Kozaki, et al.
        Liquid crystal display device comprising a \\
    & & retardation compensation layer having a maximum principal\\
    & & refractive index in the thickness direction \\

5171469 & Dec 15, 1992 &  Hittich, et al.
        Liquid-crystal matrix display \\

5175638 &  Dec 29, 1992 &  Kanemoto, et al.
        ECB type liquid crystal display device having\\
    & & birefringent layer with equal refractive indexes in the thickness\\
    & & and plane directions\\

\hline
\end{tabular}
\end{center}
\end{table}

\begin{table}
\caption{Patents on the liquid-crystal display\label{patinfoD}}
\begin{center}
\renewcommand{\arraystretch}{0.83}
\begin{tabular}{|r|r|l|}
\hline
patent  & date         & author(s) and title \\
\hline

5243451 &  Sep 7, 1993 &  Kanemoto, et al.
        DAP type liquid crystal device with cholesteric\\
    & & liquid crystal birefringent layer\\

5283677 &  Feb  1, 1994 &  Sagawa, et al.
        Liquid crystal display with ground regions \\
    & & between terminal groups\\

5308538 & May  3, 1994 &  Weber, et al.
        Supertwist liquid-crystal display \\

5319478 &  June 7, 1994  &   Funfschilling, et al.
        Light control systems with a circular polarizer\\
    & & and a twisted nematic liquid crystal having a minimum path\\
    & & difference of .lambda./2\\

5374374 & Dec 20, 1994 &  Weber, et al.
        Supertwist liquid-crystal display \\

5408346 &  Apr 18, 1995  &   Trissel, et al.
        Optical collimating device employing cholesteric\\
    & & liquid crystal and a non-transmissive reflector\\

5539578 &  Jul 23, 1996  &   Togino, et al.
        Image display apparatus\\

5543077 & Aug  6, 1996 &  Rieger, et al.
        Nematic liquid-crystal composition \\

5555116 &  Sep 10, 1996 &  Ishikawa, et al.
        Liquid crystal display having adjacent\\
    & & electrode terminals set equal in length \\

5683624 & Nov  4, 1997 &  Sekiguchi, et al.
        Liquid crystal composition \\

5771124 &  Jun 23, 1998  &   Kintz, et al.
        Compact display system with two stage magnification\\
    & & and immersed beam splitter\\

5855814 & Jan  5, 1999 &  Matsui, et al.
        Liquid crystal compositions and liquid crystal\\
    & & display elements\\

5991084 &  Nov 23, 1999 &  Hildebrand, et al.
        Compact compound magnified virtual image\\
    & & display with a reflective/transmissive optic\\

6005720 &  Dec 21, 1999 &  Watters, et al.
        Reflective micro-display system \\ \hline

\end{tabular}
\end{center}
\end{table}

But, in this network there should be thousands of 'themes'.
How to identify them?
Using the arc weights we can define a \emph{theme} as a connected
small subnetwork of size in the interval $k$ .. $K$
(for example, between $k = \frac{1}{3}h$ and $K = 3h$)
with stronger internal cohesion relatively to its
neighborhood.

To find such subnetworks we use again the arc-cuts.
We select a treshold $t$ and delete all arcs with weight
lower than $t$. In the so reduced network we determine (weakly)
connected components. The components of size in range $k .. K$,
we call them $(k,K)$-\emph{islands},
represent the themes since:
\begin{itemize}
 \item they are connected and of selected size,
 \item all arcs linking them to their outside neighbors have weight lower
       than $t$, and
 \item each vertex of an island is linked with some other
       vertex in the same island with an arc with a weight
       at least $t$.
\end{itemize}
We discard components of size smaller than $k$ as 'noninteresting'.

The components of size larger then $K$ are too large. They contain
several themes. To identify them we repeat the procedure on the
network of these components with a higher threshold value $t'$.
Recently we developed an algorithm, named \emph{Islands} \cite{Islands},
that by 'continuosly' changing the threshold identifies all maximal
$(k,K)$-islands.

\begin{figure}[!]
 \begin{center}
  \includegraphics[width=140mm,viewport=0 10 500 470,clip=]{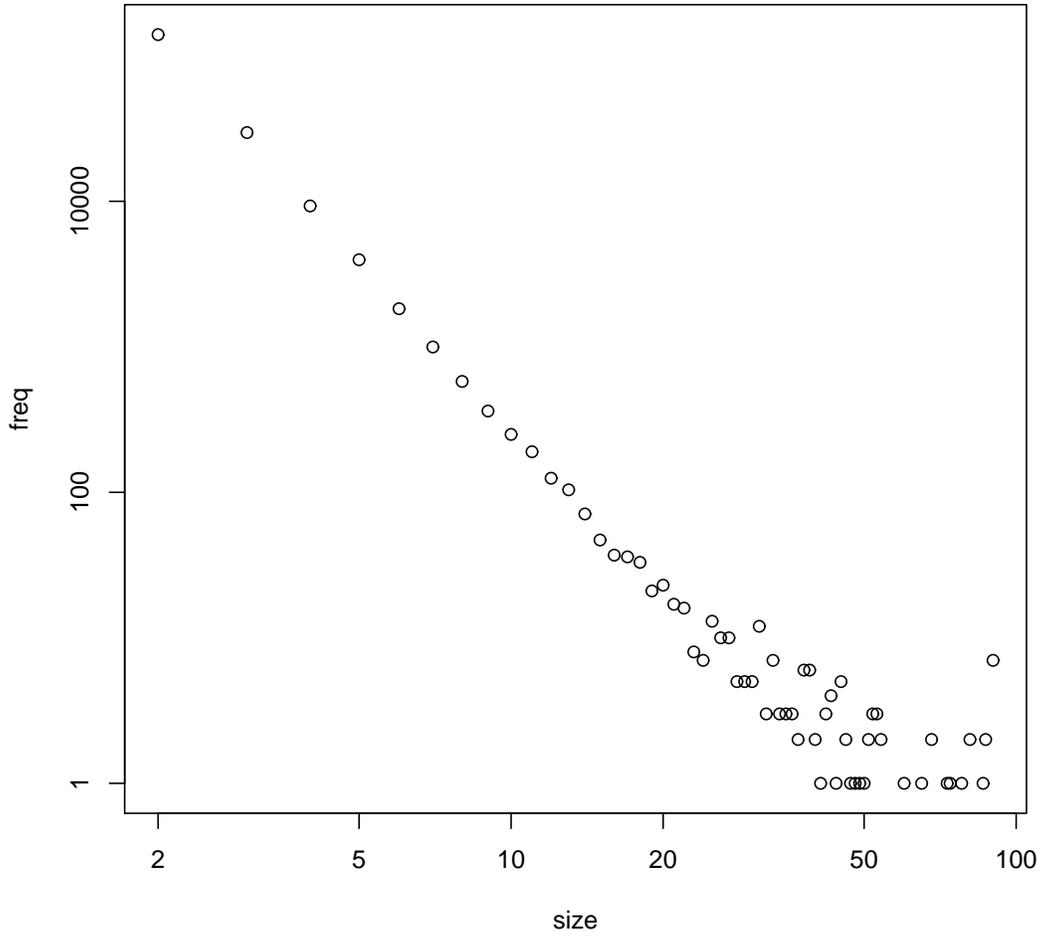}
  \caption{Island size frequency distribution \label{power}}
 \end{center}
\end{figure}

We determined for SPC weights all (2,90)-islands in the US Patents
network. The reduced network of islands has 470137 vertices, 307472 arcs and for
different $k$: $C_2 = $187610, $C_5 = $8859,$C_{30} = $101,
$C_{50} = $30 islands. The detailed island size frequency distribution
is given in Table~\ref{fris} and presented in a log-log scale
in Figure~\ref{power} that shows that it obeys the power law.

\begin{table}
\caption{Island size frequency distribution\label{fris}}
\begin{center}
{\renewcommand{\baselinestretch}{0.7}\small
\begin{verbatim}
         [1]   0 139793  29670 9288 3966 1827 997 578 362 250
        [11] 190    125    104   71   47   37  36  33  21  23
        [21]  17     16      8    7   13   10  10   5   5   5
        [31]  12      3      7    3    3    3   2   6   6   2
        [41]   1      3      4    1    5    2   1   1   1   1
        [51]   2      3      3    2    0    0   0   0   0   1
        [61]   0      0      0    0    1    0   0   2   0   0
        [71]   0      0      1    1    0    0   0   1   0   0
        [81]   2      0      0    0    0    1   2   0   0   7
\end{verbatim}
}
\end{center}
\end{table}

\begin{figure}[!]
 \begin{center}
  \includegraphics[width=160mm,viewport=45 25 620 780,clip=]{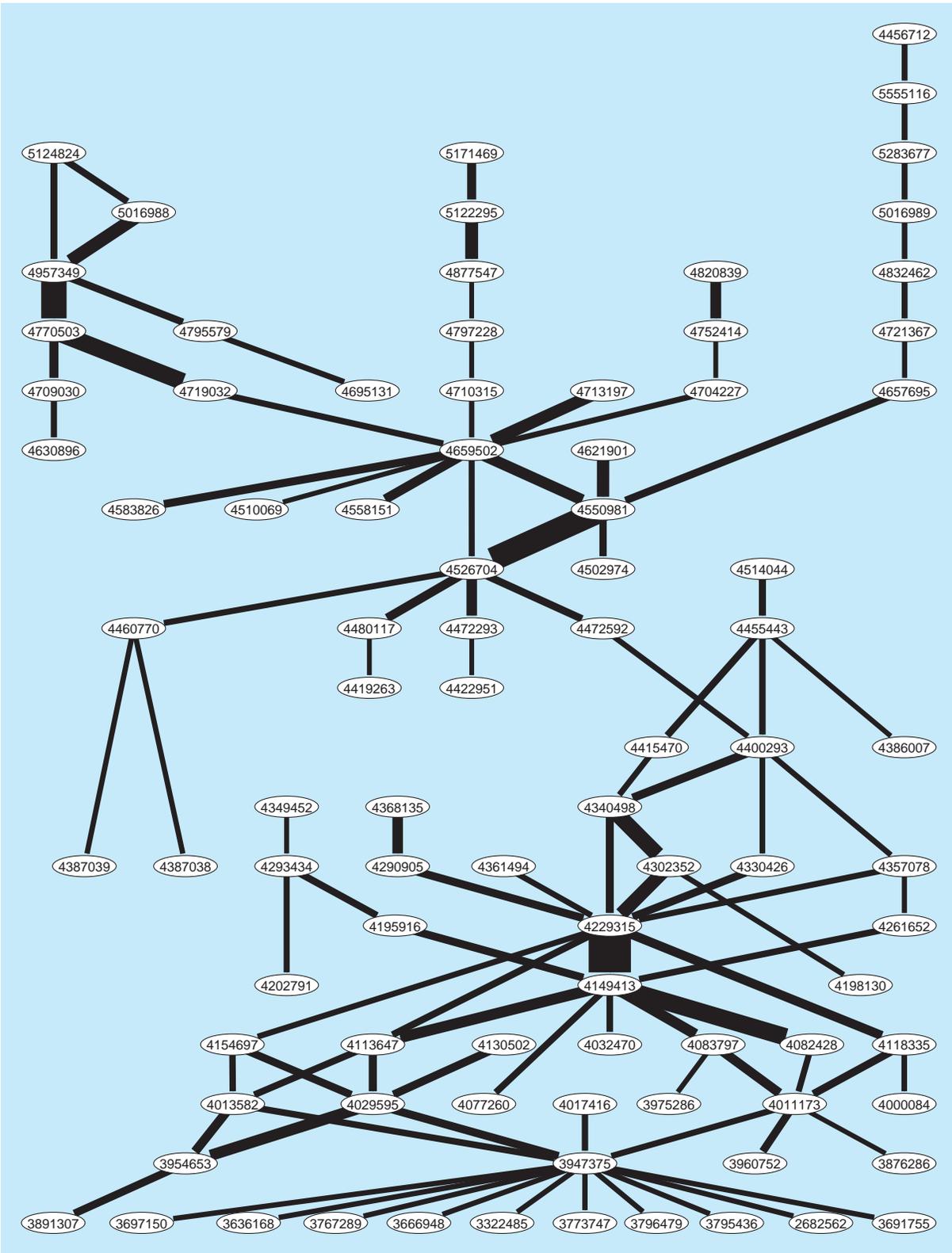}
  \caption{Main island 'liquid-crystal display'  \label{M}}
 \end{center}
\end{figure}

\begin{figure}[!]
 \begin{center}
  \includegraphics[width=160mm,viewport=65 20 660 690,clip=]{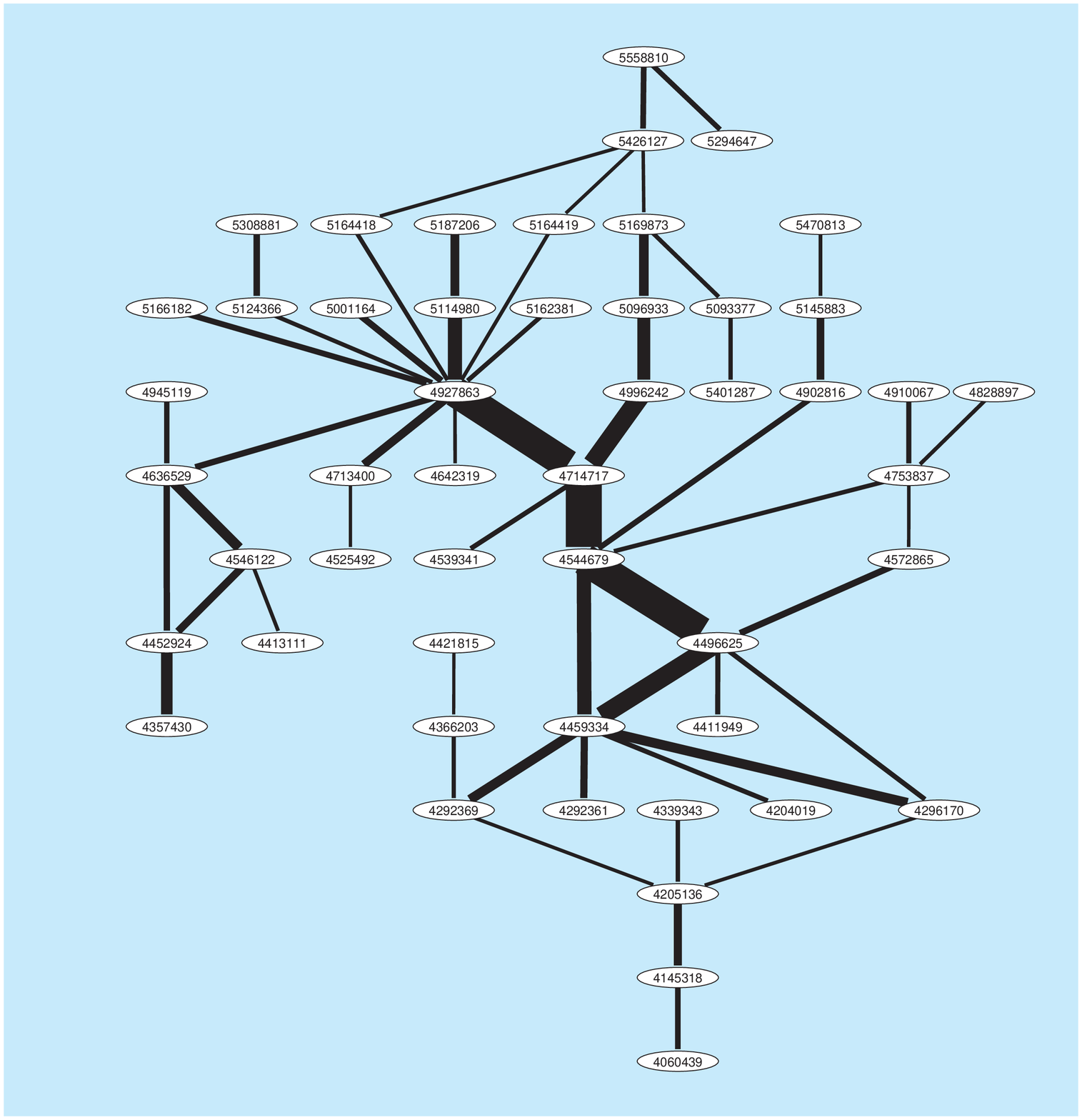}
  \caption{Island 'producing a foam' \label{A}}
 \end{center}
\end{figure}

\begin{table}
\caption{Some patents from the 'foam' island\label{Ainfo}}
\begin{center}
\begin{tabular}{|r|r|l|}
\hline
patent  & date         & author(s) and title \\
\hline
4060439 &  Nov 29, 1977 & Rosemund, et al.
           Polyurethane foam composition and method of\\
      & &  making same\\

4292369 & Sep 29, 1981 & Ohashi, et al.
        Fireproof laminates\\

4357430 &  Nov 2, 1982 &  VanCleve.
           Polymer/polyols, methods for making same and\\
      & &  polyurethanes based thereon\\

4459334 & Jul 10, 1984 & Blanpied, et al.
        Composite building panel\\

4496625 & Jan 29, 1985 &  Snider ,   et al.
        Alkoxylated aromatic amine-aromatic polyester\\
        & & polyol blend and polyisocyanurate foam therefrom\\

4544679 & Oct 1, 1985  & Tideswell, et al.
        Polyol blend and polyisocyanurate foam\\
        & & produced therefrom\\

4714717 & Dec 22, 1987 & Londrigan, et al.
        Polyester polyols modified by low molecular\\
        & & weight glycols and cellular foams therefrom\\

4927863 & May 22, 1990 & Bartlett, et al.
        Process for producing closed-cell polyurethane \\
        & & foam compositions expanded with mixtures of blowing agents\\

4996242 &  Feb 26, 1991 & Lin.
           Polyurethane foams manufactured with mixed\\
      & &  gas/liquid blowing agents\\

5169873 & Dec 8, 1992  & Behme, et al.
        Process for the manufacture of foams with the aid\\
        & & of blowing agents containing fluoroalkanes and fluorinated\\
        & & ethers, and foams obtained by this process\\

5187206 & Feb 16, 1993 & Volkert, et al.
        Production of cellular plastics by the\\
    & & polyisocyanate polyaddition process, and low-boiling,\\
    & & fluorinated or perfluorinated, tertiary alkylamines\\
    & & as blowing agent-containing emulsions for this purpose\\

5308881 & May  3, 1994 & Londrigan, et al.
        Surfactant for polyisocyanurate foams\\
    & & made with alternative blowing agents\\

5558810 &  Sep 24, 1996 & Minor, et al.
           Pentafluoropropane compositions\\

\hline
\end{tabular}
\end{center}
\end{table}

\begin{figure}[!]
 \begin{center}
  \includegraphics[width=160mm,viewport=45 20 560 680,clip=]{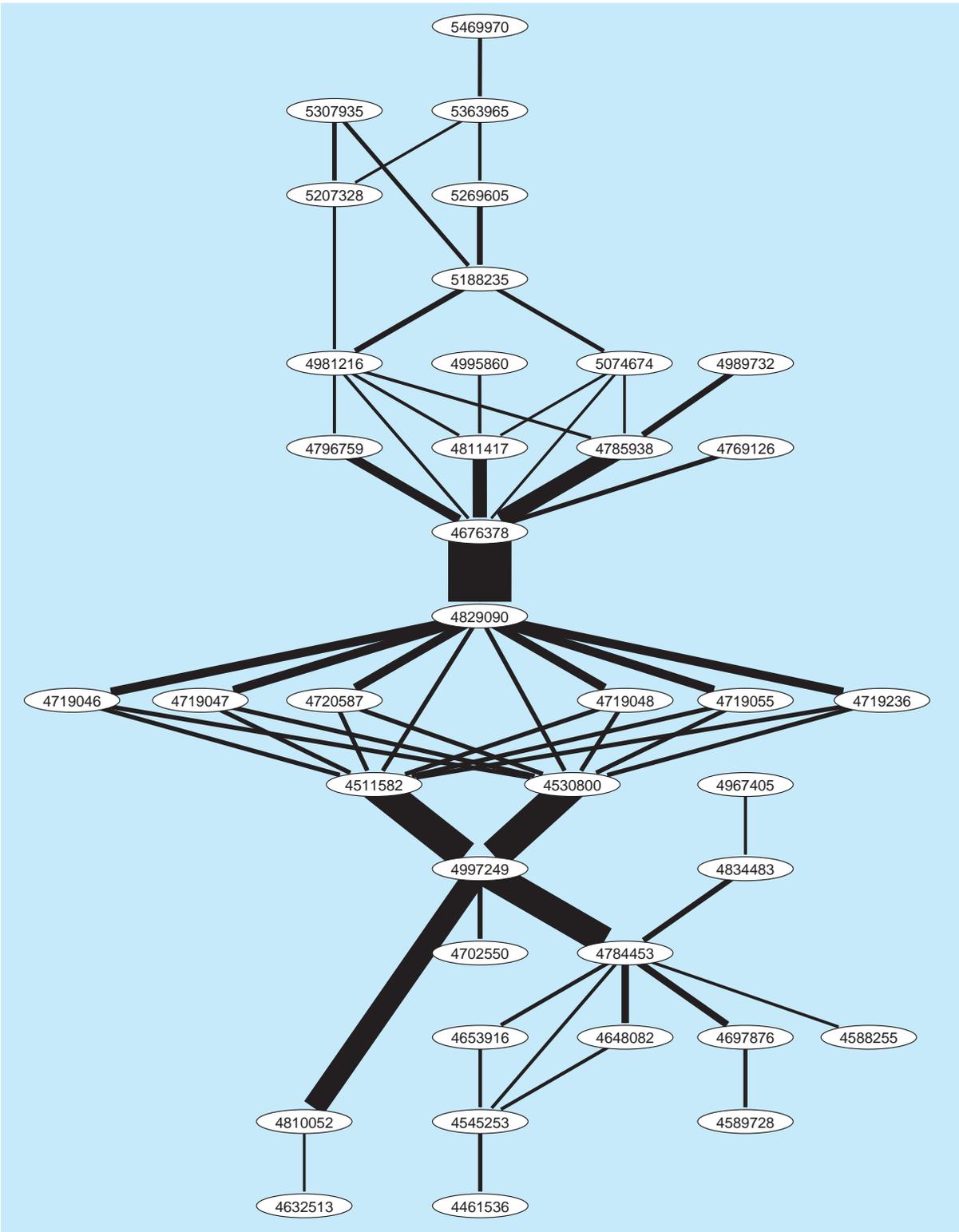}
  \caption{Island 'fiber optics and bags' \label{C}}
 \end{center}
\end{figure}

\begin{table}
\caption{Some patents from 'fiber optics and bags' island\label{Cinfo}}
\begin{center}
\begin{tabular}{|r|r|l|}
\hline
patent  & date          & author(s) and title \\
\hline
4461536 & Jul 24, 1984 & Shaw, et al.
        Fiber coupler displacement transducer\\

4511582 & Apr 16, 1985 & Bair.
        Phenanthrene derivatives\\

4530800 & Jul 23, 1985  & Bair.
        Perylene derivatives\\

4589728 & May 20, 1986 & Dyott, et al.
        Optical fiber polarizer\\

4676378 & Jun 30, 1987  & Baxley, et al.
        Bag pack\\

4719047 & Jan 12, 1988  & Bair.
        Anthracene derivatives\\

4784453 & Nov 15, 1988  & Shaw, et al.
        Backward-flow ladder architecture and method\\

4785938 & Nov 22, 1988 & Benoit, Jr., et al.
        Thermoplastic bag pack\\

4810052 & Mar  7, 1989 & Fling.
        Fiber optic bidirectional data bus tap\\

4811417 & Mar  7, 1989 & Prince, et al.
        Handled bag with supporting slits in handle\\

4829090 & May 9, 1989   & Bair.
        Chrysene derivatives\\

4981216 & Jan  1, 1991  & Wilfong, Jr.
        Easy opening bag pack and supporting rack\\
        & & system and fabricating method\\

4997249 & Mar 5, 1991   & Berry, et al.
        Variable weight fiber optic transversal filter\\

5188235 & Feb 23, 1993  & Pierce, et al.
        Bag pack\\

5307935 & May  3, 1994 & Kemanjian.
        Packs of self opening plastic bags and method of\\
        & & fabricating the same\\

5363965 & Nov 15, 1994  & Nguyen.
        Self-opening thermoplastic bag system\\

\hline
\end{tabular}
\end{center}
\end{table}

The main island has 90 vertices and contains middle parts of the main
path and the CPM path. They also have a short common part.
Again, the greedy strategy of the main path leads to a less vital branch.
Considering the basic data about the patents from
Table~\ref{patinfo}-\ref{patinfoC}, we see that also the main island
deals with 'liquid crystal displays'.

For additional illustration of results obtained by Islands
algorithm we selected two smaller islands at lower levels --
see Figure~\ref{A} (50 vertices) and Figure~\ref{C} (38 vertices).
Retreiving the basic data about some patents in these islands from
\textbf{\textit{United States Patent and Trademark Office}},
see Table~\ref{Ainfo} and Table~\ref{Cinfo}, we can label the
corresponding theme of the first island as 'producing a foam'.
The theme of the second island deals initially with 'fiber optics',
but in the upper part it switches to 'bag pack system'.

\section{Conclusions}

In the paper we proposed an approach to the analysis of citation
networks that can be used also for very large networks with millions
of vertices and arcs.

On test cases, the methods SPC, SPLC, NPPC produced almost the
same results. Since the method SPC  has additional
'nice' properties it could be considered as a 'first choice' --
but, to make a grounded recommendation,
additional experiences should be gained from the analyses of real-life
large citation networks.

The granularity of the results strongly depends on the range
for 'interesting themes' $k$ .. $K$ -- varying these two parameters
we get larger or smaller sets of themes.

Instead of arc-cuts we could consider also vertex-cuts
with respect to $p$-cores on SPC weights \cite{pCores}
with a $p$-function
\[ p(v,W) = \max( \sum_{u \in W : u R v} w(u,v),
                  \sum_{u \in W : v R u} w(v,u) ) \]

The subnetworks approach only filters out the structurally important
subnetworks thus providing a researcher with a smaller manageable
structures which can be further analyzed using more sophisticated
and/or substantial methods.

\section{Acknowledgments}

The search path count algorithm was developed during my visit in
Pittsburgh in 1991 and presented at the Network seminar
\cite{Bat91}. It was presented to the broader audience
at EASST'94 in Budapest \cite{Bat94}. In 1997 it was
included in program \texttt{\textbf{Pajek}} \cite{pajek}.
The 'preprint' transformation was developed as a part of the
contribution for the Graph drawing contest 2001 \cite{GD01}.
The algorithm for the path length counts was developed in August 2002
and the Islands algorithm in August 2003.

The author would like to thank Patrick Doreian and Norm Hummon
for introducing him into the field of citation network analysis,
Eugene Garfield for making available the data on real-life
networks and providing some relevant references,
and Andrej Mrvar and Matja\v{z} Zaver\v{s}nik
for implementing the algorithms in  \texttt{\textbf{Pajek}}.

This work was supported by the Ministry of Education, Science and Sport of
Slovenia, Project 0512-0101.


\newpage

\begin{thebibliography}{99}

\bibitem{Asimov}  Asimov I.: The Genetic Code,
  New American Library, New York, 1963.

\bibitem{Bat91} Batagelj V.: Some Mathematics of Network Analysis.
 Network Seminar, Department of Sociology,
 University of Pittsburgh, January 21, 1991.

\bibitem{Bat94} Batagelj V.: An Efficient Algorithm for Citation Networks
 Analysis. Paper presented at EASST'94, Budapest, Hungary,
 August 28-31, 1994.

\bibitem{pajek} Batagelj V., Mrvar A.: \texttt{\textbf{Pajek}} -- program for
 analysis and visualization of large networks. \\
 \url{http://vlado.fmf.uni-lj.si/pub/networks/pajek/}\\
 \url{http://vlado.fmf.uni-lj.si/pub/networks/pajek/howto/extreme.htm}

\bibitem{GD01} Batagelj V., Mrvar A.:
 Graph Drawing Contest 2001 Layouts \\
 \url{http://vlado.fmf.uni-lj.si/pub/GD/GD01.htm}

\bibitem{pCores} Batagelj V., Zaver\v{s}nik M.:
 Generalized Cores.  Submitted, 2002.\\
 \url{http://arxiv.org/abs/cs.DS/0202039}

\bibitem{Islands} Batagelj V., Zaver\v{s}nik M.:
 Islands -- identifying themes in large networks. In preparation, August 2003.

\bibitem{BW} Brandes U., Willhalm T.:
 Visualization of
 bibliographic networks with a reshaped landscape metaphor.
 Joint Eurographics -- IEEE TCVG Symposium on Visualization,
 D. Ebert, P. Brunet, I. Navazo (Editors), 2002.\\
 \url{http://algo.fmi.uni-passau.de/\symbol{126}brandes/}\\
 \url{\strut\qquad publications/bw-vbnrl-02.pdf}

\bibitem{algo}  Cormen T.H.,  Leiserson C.E., Rivest R.L.,  Stein C.:
 Introduction to Algorithms, Second Edition. MIT Press, 2001.

\bibitem{Gar64} Garfield E, Sher IH, and Torpie RJ.:
 The Use of Citation Data in Writing the History of Science.
 Philadelphia: The Institute for Scientific Information, December 1964.\\
\url{http://www.garfield.library.upenn.edu/papers/}\\
\url{\strut\qquad useofcitdatawritinghistofsci.pdf}

\bibitem{Gar01} Garfield E.:
 From Computational Linguistics to Algorithmic Historiography,
 paper presented at the Symposium in Honor of Casimir Borkowski
 at the University of Pittsburgh School of Information Sciences,
 September 19, 2001.\\
 \url{http://garfield.library.upenn.edu/papers/pittsburgh92001.pdf}

\bibitem{Gar02} Garfield E., Pudovkin A.I.,  Istomin, V.S.:
 \textit{\textbf{Histcomp}} -- (\textit{comp}iled \textit{Hist}oriography program)\\
 \url{http://garfield.library.upenn.edu/histcomp/guide.html}\\
 \url{http://www.garfield.library.upenn.edu/histcomp/index.html}

\bibitem{3D} Garner R.:
 A computer oriented, graph theoretic analysis of citation index structures.
 Flood B. (Editor), Three Drexel information science studies, Philadelphia:
 Drexel University Press 1967.\\
 \url{http://www.garfield.library.upenn.edu/rgarner.pdf}

\bibitem{HumDor89} Hummon N.P., Doreian P.:
 Connectivity in a Citation Network: The Development of DNA Theory.
 Social Networks, {\bf 11}(1989) 39-63.

\bibitem{HumDor90} Hummon N.P., Doreian P.:
 Computational Methods for Social Network Analysis.
 Social Networks, {\bf 12}(1990) 273-288.

\bibitem{HuDoFr90} Hummon N.P., Doreian P., Freeman L.C.:
 Analyzing the Structure of the Centrality-Productivity Literature
 Created Between 1948 and 1979.
 Knowledge: Creation, Diffusion, Utilization, {\bf 11}(1990)4, 459-480.

\bibitem{ha} Kleinberg J.:
 Authoritative sources in a hyperlinked environment.
 In Proc 9th ACMSIAM Symposium on Discrete Algorithms, 1998, p. 668-677.\\
 \url{http://www.cs.cornell.edu/home/kleinber/auth.ps}\\
 \url{http://citeseer.nj.nec.com/kleinberg97authoritative.html}

\bibitem{GT} Wilson, R.J., Watkins, J.J.:
 \emph{Graphs: An Introductory Approach}.
 New York: John Wiley and Sons, 1990.

\bibitem{data} Pajek's datasets -- citation networks:\\
 \url{http://vlado.fmf.uni-lj.si/pub/networks/data/cite/}

\bibitem{HEP} KDD Cup 2003:\\
 \url{http://www.cs.cornell.edu/projects/kddcup/index.html}\\
 \url{http://arxiv.org/}

\bibitem{patents} Hall, B.H., Jaffe, A.B. and Tratjenberg M.:
 The NBER U.S. Patent Citations Data File. NBER Working Paper 8498 (2001).\\
 \url{http://www.nber.org/patents/}

\bibitem{uspto} The United States Patent and Trademark Office. \\
 \url{http://patft.uspto.gov/netahtml/srchnum.htm}

\bibitem{SOMLVQ}
 Bibliography on the Self-Organizing Map (SOM) and Learning Vector Quantization (LVQ)\\
 \url{http://liinwww.ira.uka.de/bibliography/Neural/SOM.LVQ.html}

\bibitem{SOM}
 Neural Networks Research Centre: Bibliography of SOM papers.\\
 \url{http://www.cis.hut.fi/research/refs/}

\end{thebibliography}
\end{document}